\documentclass[letterpaper,twocolumn,pra,
aps,showpacs,superscriptaddress,floatfix]{revtex4-2}

\usepackage[latin1]{inputenc}
\usepackage{bbm}
\usepackage{subcaption}
\usepackage{bm}
\usepackage[usenames]{color}
\usepackage{multirow}
\usepackage{amssymb}
\usepackage{amsbsy}
\usepackage{mathtools}
\usepackage{amsmath}
\usepackage{stmaryrd}
\usepackage{graphicx}
\usepackage{epsfig}
\usepackage{placeins}
\usepackage{bbold}
\usepackage{braket}
\usepackage{blindtext}
\usepackage[colorlinks,linkcolor=blue,citecolor=blue,urlcolor=blue]{hyperref}

\newcommand{\be}{\begin{equation}}

\newcommand{\ee}{\end{equation}}

\makeatletter

\begin{document}
	\title{Interplay of decoherence and relaxation in a two-level system interacting with an infinite-temperature reservoir}
	
	
	\author{Jiaozi Wang}
	\email{jiaozi.wang@uos.de}
\affiliation{Department of Mathematics/Computer Science/Physics, University of Osnabr\"uck, D-49076 
Osnabr\"uck, Germany}

	\author{Jochen Gemmer}
	\email{jgemmer@uos.de}
\affiliation{Department of Mathematics/Computer Science/Physics, University of Osnabr\"uck, D-49076 
Osnabr\"uck, Germany}

	\begin{abstract}
	We study the time evolution of a single qubit in contact with a bath, within the framework of projection operator methods. Employing the so-called modified Redfield theory which also treats energy conserving interactions non-perturbatively, we are able to study the regime beyond the scope of the ordinary approach. Reduced equations of motion for the qubit are derived in a idealistic system where both the bath and system-bath interactions are modeled by Gaussian distributed random matrices. In the strong decoherence regime, a simple relation between the bath correlation function and the decoherence process induced by the energy conserving interaction is found. It implies that energy conserving interactions slow down the relaxation process, which leads to a zeno freezing if they are sufficiently strong. Furthermore, our results are also confirmed in numerical simulations.

	\end{abstract}
	
	\maketitle
\section{Introduction}

In the field of open quantum systems,  the question of whether or how a small quantum system evolves to a steady state, when being coupled to a large bath, has attracted significant attention and been studied extensively in recent decades in various fields of physics \cite{breuer2002book,joos2013book,weiss2012book, WGW08,Daniel05,fialko2014decoherence,gemmer2006,gemmer2007projection,Jin13,Maxim14,Raedt17,santra2017,yuan2011decoherence,gorin2006dynamics}.

On the route of the system evolving towards equilibrium state, decoherence and relaxation are two fundamental processes which often coexist and may in general be correlated to each other. It is natural to ask in which exact way decoherence and relaxation processes are related. Or more specifically, as focused on in this paper: what is the impact of decoherence on the relaxation process?
The question has been discussed in the weak coupling regime \cite{esposito2005emergence}, as well as in case of the pure-dephasing interaction \cite{Lars_PhysRevA.99.012118,vznidarivc2013transport}. 
The effect of spatial decoherence on the transport properties of particle(s) are investigated in ordered \cite{esposito2005emergence}
and disordered \cite{vznidarivc2013transport} tight-binding lattices. 
In Ref. \cite{Lars_PhysRevA.99.012118}, based on the memory kernel approach, the dynamics of the system is found to be slowed down by decoherence, which leads to the transition towards zeno freezing \cite{misra1977zeno} in the
strong decoherence regime.
However, not so much is known for more generic spin-bath coupling.

The answer to this question relies on the knowledge of the time evolution of reduced density matrix (RDM) of the system of interest.
However, for most open
quantum systems, which are not exactly solvable, it is too complicated to obtain it
without making approximations.
The standard procedure in
theoretical treatments is to derive closed approximate equations of motion of the system, the quantum master equation (QMEQ) \cite{breuer2002book,lidar2019lecture,weiss2012quantum}, from the
underlying time-dependent Schr\"{o}dinger equation (TDSE) by
eliminating the environmental degrees of freedom.

One of the most important and commonly used methods to derive the QMEQ is the projection method (such as Nakajima-Zwanzig method \cite{nakajima1958quantum,zwanzig1960ensemble} or the
time-convolutionless method \cite{shibata1977generalized,chaturvedi1979time,shibata1980expansion,uchiyama1999unified}). 
Another foundational method is the Hilbert space average method \cite{gemmer2007projection}, which estimates conditional
quantum expectation values through an appropriate average over a constraint region in Hilbert space.
Besides the usual master equation approach, there are also other methods aiming to derive closed equations of the system, e.g. approach based on resonance theory \cite{merkli2007decoherence,merkli2008dynamics}, linear-response theory \cite{carrera2014single}, and Dyson Brownian technique \cite{genway2013dynamics}. 

 For weak system-bath coupling, the Redfield equation  \cite{redfield1957theory} can be derived by  keeping the perturbations to second order and making use of the Born-Markov approximation. In this case, exact equations for relaxation and decoherence process can be derived. To understand the relation between these two processes better, it is desirable to go beyond the weak system-bath coupling regime, which is always a challenging task. 
One possible method to achieve this is to straightforwardly extent the perturbation theory to higher orders. Another possible method is to treat a certain part of the interaction Hamiltonian non-perturbatively, as done e.g., in the F\"{o}rster theory \cite{forster1946energiewanderung,forster1948zwischenmolekulare} and the modified Redfield theory \cite{zhang1998exciton}.  

In our paper, we study the problem in a idealistic system where a single qubit is coupled to a bath modeled by random matrix \cite{esposito2003quantum, gemmer2006,carrera2014single,genway2013dynamics} via system-bath interaction consisting of both energy-conserving  (\textit{EC}) and energy-exchange (\textit{EX}) parts (with respect to system energy).
We employ the modified Redfield theory \cite{zhang1998exciton,trushechkin2019calculation,trushechkin2022quantum,seibt2017ultrafast,yang2002influence}, where the \textit{EC} interaction is also treated  non-perturbatively. Reduced equations of motion for the system in the interaction picture is derived, which are valid for arbitrary large \textit{EC} interaction whenever the \textit{EX} interaction is weak compared to the unperturbed Hamiltonian. Employing further assumptions, we also derive the equations for the time evolution of system's RDM in the Schr\"{o}dinger picture. A simple relation is found between the bath correlation function and the decoherence process induced by the \textit{EC} interaction, which implies that
relaxation process is slowed down by the \textit{EC} interaction, leading to a zeno freezing if it is sufficiently strong.
The paper is organized as follows.
In Sec. \ref{sect-setup}, we introduce the general setup and in Sec. \ref{sect-RF} the reduced equations of motion for the system are derived.  In Sec. \ref{sect-GORM}, we apply our analytical results to a spin random matrix-model, while the results are checked numerically in Sec. \ref{sect-numerical}. Conclusions and outlook are given in Sec. \ref{Sect-Con}

\section{General Setup}\label{sect-setup}

We consider a model where a single qubit is coupled to a bath, the Hamiltonian of which reads,
\begin{gather}
H=H_{\text S}+H_{\text B}+V, \nonumber \\ 
V = \sum_{m,n=1}^{2}\lambda_{mn}|m\rangle\langle n|\otimes H_{mn},
\label{eq-HSimple}
\end{gather}
where $H_{\text S} = \omega\sigma^{\text S}_z$ ($\sigma^{\text S}_{z}$ denotes the $z$-direction Pauli operator of the system), $H_{\text B}$ and $V$ indicate system, bath and interaction Hamiltonian, respectively.
Eigenstates of $H_{\text S}$ and $H_{\text B}$ are denoted by $|m\rangle$ and $|E_k\rangle$, 
\be
H_{\text S} |m\rangle = e_m |m\rangle,\quad H_{\text B}|E_k\rangle = E_k |E_k\rangle,
\ee
where $m=1,2$ correspond to spin down and up.
$H_{mn}$ are some generic operators in the Hilbert space of the bath satisfying
\be
\left\Vert H_{mn}\right\Vert =\left\Vert H_{\text B}\right\Vert.
\ee
$\Vert \bullet \Vert$ indicates the norm of the operator which is defined as (for a generic operator $\cal O$)
\be
\Vert {\cal O} \Vert \equiv \text{Tr}\{{\cal O}\} {\cal O}^\dagger).
\ee

The Hermiticity of the interaction Hamiltion requires 
\be
(\lambda_{mn}H_{mn})^{\dagger}=\lambda_{nm}H_{nm}
\ee
holds for $m,n=1,2$.
The interaction Hamiltonian can be divided into a energy-conserving (\textit{EC}) part (denoted by $V_{\text{ec}}$) and a energy-exchange (\textit{EX}) part (denoted by $V_{\text{ex}}$)
\be
V = V_{\text{ec}} + V_{\text{ex}},
\ee
where
\begin{align}
V_{\text{ec}} & =\lambda_{11}|1\rangle\langle1|\otimes H_{11}+\lambda_{22}|2\rangle\langle2|\otimes H_{22}, \nonumber \\
V_{\text{ex}} & =\lambda_{12}|1\rangle\langle2|\otimes H_{12}+\lambda_{12}^{*}|2\rangle\langle1|\otimes H_{12}^{\dagger}.
\end{align}
The initial state 
considered here is a product state, written as
\be\label{eq-initial}
\rho(0)=\rho^{\text S}(0)\otimes \rho^{\text B} (0),
\ee
where $\rho^{\text S}(0) = |\psi_{\text S}(0)\rangle\langle \psi_{\text S}(0)|$ and 
\be
\rho^{\text B}(0)=\frac{1}{Z}\exp(-\beta H_{B}),\quad Z=\text{Tr}\{ \exp(-\beta H_{B})\}.
\ee
In our paper we only consider the simplest case where the bath is at infinite temperature $\beta = 0$. $\rho^\text{B}(0)$ reads
\be
\rho^\text{B}(0) = \frac{\mathbb{1}_\text{B}}{d},
\ee
where $d$ denotes the Hilbert space dimension of the bath.
To study time evolution of the RDM of the system $\rho^{\text S}(t)=\text{Tr}_B(\rho(t))$, we divide the Hamiltonian into an unperturbed part $H_0$ and a perturbation part $H^{\text{int}}$,
\be
H = H_0 + H^{\text{int}}.
\ee
As the key ingredient in the modified Redfield theory \cite{zhang1998exciton}, the unperturbed Hamiltonian 
\be\label{eq-H0}
H_{0}=H_{\text S}+H_{\text B}+V_{\text{ec}},
\ee
also comprises the \textit{EC} interaction
$V_{\text{ec}}$,
while the perturbation 
\be
H^{\text{int}} = V_{\text{ex}}
\ee
only consists of the \textit{EX} interaction.
In our paper, we only consider the situation where $\Vert H^{\text{int}}\Vert \ll \Vert H_0 \Vert$. In this case system's eigenbasis $|m\rangle$ usually forms a good preferred basis   \cite{he2014statistically,paz1999quantum,WGW08}, or e.g., the stationary RDM is approximately diagonal in $|m\rangle$. Thus in the rest of the paper, whenever talking about decoherence, we refer to decoherence in the eigenbasis of the system.

In the interaction picture, the density matrix of the composite system at time $t$ is written as
\be\label{eq-rhoI}
\rho_{I}(t)=\exp(iH_{0}t)\rho(t)\exp(-iH_{0}t),
\ee
where
\begin{align}
\exp(-iH_{0}t) & =|m\rangle\langle m| \otimes U_m(t), \label{eq-U0} \\
U_m(t) & \equiv \exp(-i(H_{\text B}^{(m)}+e_{m})t).
\end{align}
$H_{\text B}^{(m)}$ can be regarded as an effective bath Hamiltonian with respect to system state $|m\rangle$, defined as
\be\label{eq-HBM}
H_{\text B}^{(m)}=H_{\text B}+\lambda_{mm}H_{mm}.
\ee
The spectral density of $H_{\text B}^{(m)}$ is denoted by $\Omega_m(E)$, which will be used later in Sec. \ref{sect-GORM} .
Denoting the eigenvalues and eigenstates of $H^{(m)}_{\text B}$ by $E^{(m)}_k$ and $|E^{(m)}_k\rangle$, one gets
\be
H_{0}|m\rangle|E_{k}^{(m)}\rangle=(E_{k}^{(m)}+e_{m})|m\rangle|E_{k}^{(m)}\rangle.
\ee
Similarly, one has
\begin{align}
H^{\text{int}}_{I}(t) = & \exp(iH_{0}t)H^{\text{int}}_{I}\exp(-iH_{0}t), \nonumber \\
= & \lambda_{12}|1\rangle\langle2|\otimes U_{1}^{\dagger}(t)H_{12}U_{2}(t)+h.c..\label{eq-VI}
\end{align}
The RDM of the qubit in the interaction picture can be written as
\be
\rho_{I}^{\text S}(t)  = \text{Tr}_{\text B}\{\rho_{I}(t)\}, 
\ee
where the matrix elements
$[\rho_{I}^{\text S}(t)]_{mn}=\langle m|\rho_{I}^{\text S}(t)|n\rangle$ can be written as
\be
[\rho_{I}^{\text S}(t)]_{mn}=\begin{cases}
\rho_{mn}^{\text S}(t) & m=n\\
\langle m|\text{Tr}_{\text B}\{U_{m}^{\dagger}(t)\rho(t)U_{n}(t)\}|n\rangle & m\neq n
\end{cases}.\label{eq-rhoI}
\ee
Here $\rho_{mn}^{\text S}(t)\equiv\langle m|\rho^{\text S}(t)|n\rangle$ indicate the matrix elements of RDM in the Schr\"{o}dinger picture.
One can see that, in the  modified Redfield approach,  the diagonal elements of RDM in the interaction picture are the same as in the Schr\"{o}dinger picture. Note that due to  $[H_{0},H_{\text{S}}+H_{\text{B}}]\neq0$, which is different from the usual approaches, there is no simple relation between off-diagonal elements in the interaction and Schr\"{o}dinger  picture. This is known to be a main drawback of this method,  as the decoherence dynamics in the Schr\"{o}dinger picture can not be straightforwardly studied \cite{trushechkin2019calculation,yang2002influence, seibt2017ultrafast,valkunas2013molecular,ishizaki2009unified,novoderezhkin2010physical}.  In this paper, we will tackle this problem by employing further assumptions.

\section{Reduced equations of motion}\label{sect-RF}
In this section, we are going to derive the reduced equations of motion for the system in both interaction and Schr\"{o}dinger picture.
%
\subsection{Time evolution of the system in the interaction picture}
In this section, we derive reduced equations of motion for the system in the interaction picture using modified Redfield theory. Different from the derivations shown in Refs. \cite{yang2002influence,seibt2017ultrafast} which only consider the diagonal elements of the RDM of the system, we also study the time evolution of off-diagonal elements of RDM as well. Here we only focus on the simplest case where the interaction Hamiltonians $H_{mn}$ are uncorrelated with each other.
As a guideline of this section, we would like to mention here, that the derivations given below are almost the same as the standard derivations of the quantum master equations based on the Born-Markov approach, but with a different unperturbed Hamiltonian given in Eq. \eqref{eq-H0}.   More detailed derivations can be found in Appendix A.

Let's consider a projection superoperator $\cal P$, defined as
\be
{\cal P}\rho=\text{Tr}_{\text B}\{\rho\}\otimes \frac{\mathbbm{1}_B}{d}.
\ee
Applying $\cal P$ to the density matrix in the interaction picture yields
\be
{\cal P}\rho_{I}(t)=\text{Tr}_{\text B}\{\rho_{I}(t)\}\otimes \frac{\mathbbm{1}_{\text B}}{d}=\rho_{I}^{\text S}(t)\otimes \frac{\mathbbm{1}_{\text B}}{d}.
\ee
Using the general method of projection operator technique, and keeping perturbation terms up to second order, for initial states satisfying ${\cal P}\rho_I(0) = \rho_I(0)$
one has 
\be\label{eq-drho0}
\frac{{\text d}}{{\text d}t}\rho_{I}^{\text S}(t)=-\frac{1}{d}\int_{0}^{t}{\text d}s\text{Tr}_{\text B}\left\{ [H^{\text{int}}_{I}(t),[H^{\text{int}}_{I}(s),\rho_{I}^{\text S}(s)\otimes \mathbbm{1}_\mathrm{B}]]\right\}.
\ee

To study the right hand side of Eq. \eqref{eq-drho0}, it is useful to introduce a bath correlation function $F_{mn}(\tau)$ defined as
\be\label{def-Fmn}
F_{mn}(\tau)=\frac{1}{d}\text{Tr}_{\text B}\left\{ U_{m}^{\dagger}(\tau)H_{mn}U_{n}(\tau)H^\dagger_{mn}\right\}.
\ee
With straightforward derivations (see Appendix A for details), by employing the Markovian approximation, in case that the interaction Hamiltonian $H_{mn}$ are uncorrelated with each other, the time evolution of the RDM of the system can be written as, 
\begin{align}\label{eq-dRho-both}
\frac{{\text{d}}[\rho_{I}^{\text S}(t)]_{11}}{{\text{d}}t} & =-\Gamma_{r}([\rho_{I}^{\text S}(t)]_{11}-[\rho_{I}^{\text S}(\text{eq})]_{11}), \nonumber \\
\frac{\text{d}[\rho_{I}^{\text S}(t)]_{22}}{\text{d}t} & =-\Gamma_{r}([\rho_{I}^{\text S}(t)]_{22}-[\rho_{I}^{\text S}(\text{eq})]_{22}), \nonumber \\
\frac{{\text{d}}[\rho_{I}^{\text S}(t)]_{21}}{{\text{d}}t}&=-\frac{\Gamma_{r}}{2}[\rho_{I}^{\text S}(t)]_{21},
\end{align}
where 
\be
[\rho_{I}^{\text S}(\text{eq})]_{11} = [\rho_{I}^{\text S}(\text{eq})]_{22}=\frac{1}{2}.
\ee 
$\Gamma_r$ indicates the relaxation rate, which can be written as
\begin{align}
\Gamma_{r}& =4|\lambda_{12}|^{2}\int_{0}^{\infty}\mathrm{Re}[F_{12}(\tau)]{\text d}\tau
\nonumber \\
& =2|\lambda_{12}|^{2}\int_{-\infty}^{\infty}F_{12}(\tau){\text d}\tau. \label{eq-Gamma2}
\end{align}

It should be mentioned here that, the Markovian approximation can only be applied under the condition that the correlation function $F_{12}(\tau)$ decays sufficiently fast on a time $\tau_B$ (correlation time) compared to the relaxation time of the system $\tau_R$, that is, 
\be\label{eq-condition-0}
\tau_B \ll \tau_R.
\ee
From Eq. \eqref{eq-dRho-both}, one obtains the solution of reduced equations of motion of the system in the interaction picture,
\begin{align}
[\rho_{I}^{\text S}(t)]_{11}-[\rho_{I}^{\text S}(\text{eq})]_{11}&=e^{-\Gamma_{r}t}([\rho_{I}^{\text S}(0)]_{11}-[\rho_{I}^{\text S}(\text{eq})]_{11}), \nonumber \\
[\rho_{I}^{\text S}(t)]_{22}-[\rho_{I}^{\text S}(\text{eq})]_{22}&=e^{-\Gamma_{r}t}([\rho_{I}^{\text S}(0)]_{22}-[\rho_{I}^{\text S}(\text{eq})]_{22}), \nonumber \\
[\rho_{I}^{\text S}(t)]_{21}&=e^{-\frac{\Gamma_{r}}{2}t}[\rho_{I}^{\text S}(0)]_{21}.\label{eq-result-rdm}
\end{align}
One can see that, in the interaction picture, the resulting equation of motion for the diagonal and off-diagonal elements of RDM are decoupled, which is due to the second order approximation used in Eq. \eqref{eq-drho0}.
\subsection{Time evolution of RDM in the Schr\"{o}dinger picture}\label{sect-RDMS}
After having derived the reduced equations of motion for the system in the interaction picture, we continue to consider the Schr\"{o}dinger picture.
As has already been shown in Eq. \eqref{eq-rhoI}, the diagonal elements of the RDM in the Schr\"{o}dinger picture are the same as in the interaction picture, so we only need to study the off-diagonal elements. 

To this end, instead of the mixed state in Eq. \eqref{eq-initial}, it is more convenient to consider a pure state $|\psi(0)\rangle$ as an initial state, where
\be
|\psi(0)\rangle = |\psi^{\text S}(0)\rangle \otimes |\psi^{\text B}(0)\rangle.
\ee
The bath initial state is written as
\be\label{eq-initial-pure}
|\psi^{\text B}(0)\rangle=\sum_{k}c_{k}|E_{k}\rangle,
\ee
where $c_k$ are complex numbers, the real and imaginary parts of which are drawn independently from a Gaussian distribution.
Based on the idea of dynamical quantum typicality \cite{sugiura2012typ,elsayed2013typ,steinigeweg2014typ,Gemmer09}, the dynamics of the system starting from the pure initial state employed in Eq. \eqref{eq-initial-pure} is almost the same as that of the mixed initial state in Eq. \eqref{eq-initial}, if the dimension of the bath Hilbert space is large enough.
At time $t$ the state of the composite system in the interaction picture can always be written in the following form,
\be
|\psi_I(t)\rangle=|1\rangle|\psi^{1}_I(t)\rangle+|2\rangle|\psi^{2}_I(t)\rangle.
\ee
The RDM of the system in the interaction picture can then be written as
\be
[\rho_I^{\text S}(t)]_{mn}=\langle\psi_I^{n}(t)|\psi_I^{m}(t)\rangle.
\ee
Now we switch to the Schr\"{o}dinger picture. At time $t$ one has
\be
|\psi(t)\rangle=e^{-iH_{0}t}|\psi_{I}(t)\rangle=|1\rangle|\psi^{1}(t)\rangle+|2\rangle|\psi^{2}(t)\rangle,
\ee
where 
\be
|\psi^{m}(t)\rangle=e^{-i(H_{\text B}^{(m)}+e_{m})t}|\psi_{I}^{m}(t)\rangle.
\ee
Then, the RDM of the system in the Schr\"{o}dinger picture 
can be written by making use of $|\psi^m_I(t)\rangle$ as
\begin{gather}
[\rho^{\text S}(t)]_{mn}=\langle\psi_{I}^{n}(t)|e^{i(H_{\text B}^{(n)}+e_{n})t}e^{-i(H_{\text B}^{(m)}+e_{m})t}|\psi_{I}^{m}(t)\rangle.
\end{gather}

As we consider a two level system here, there is only one independent term in the off-diagonal elements
\be\label{eq-rho21-00}
[\rho^{\text S}(t)]_{21}=e^{-2i\omega t}\langle\psi_{I}^{1}(t)|e^{iH_{\text B}^{(1)}t}e^{-iH_{\text B}^{(2)}t}|\psi_{I}^{2}(t)\rangle.
\ee
At any time $t$, one can always divide  $|\psi^2_I(t)\rangle$ into a branch which is  parallel to $|\psi^1_I(t)\rangle$ and the other which is vertical to  $|\psi^1_I(t)\rangle$, as
\be\label{eq-2branch}
|\psi_{I}^{2}(t)\rangle=c(t)|\psi_{I}^{1}(t)\rangle+|\psi_{I}^{1\perp}(t)\rangle,
\ee
where
\be\label{def-c}
c(t)=\frac{\langle\psi_{I}^{1}(t)|\psi_{I}^{2}(t)\rangle}{\langle\psi_{I}^{1}(t)|\psi_{I}^{1}(t)\rangle}=\frac{[\rho_{I}^{\text S}(t)]_{21}}{\langle\psi_{I}^{1}(t)|\psi_{I}^{1}(t)\rangle}=\frac{[\rho_{I}^{\text S}(t)]_{21}}{[\rho_{I}^{\text S}(t)]_{11}}.
\ee
Inserting Eqs. \eqref{eq-2branch} and \eqref{def-c} to Eq. \eqref{eq-rho21-00}, one gets
\begin{align}
[\rho^{\text S}(t)]_{21}=&\frac{[\rho_{I}^{\text S}(t)]_{21}e^{-2i\omega t}}{\langle\psi_{I}^{1}(t)|\psi_{I}^{1}(t)\rangle}\langle\psi_{I}^{1}(t)|e^{iH_{\text B}^{(1)}t}e^{-iH_{\text B}^{(2)}t}|\psi_{I}^{1}(t)\rangle \nonumber \\
& + e^{-2i\omega t}\langle\psi_{I}^{1}(t)|e^{iH_{\text B}^{(1)}t}e^{-iH_{\text B}^{(2)}t}|\psi_{I}^{1\perp}(t)\rangle.
\end{align}
If we make the assumption that
\be\label{eq-asm-dec1}
|\langle\psi_{I}^{1}(t)|e^{iH_{\text B}^{(1)}t}e^{-iH_{\text B}^{(2)}t}|\psi_{I}^{1\perp}(t)\rangle|\approx0,
\ee
$[\rho^{\text S}(t)]_{21}$ can be related to $[\rho_I^{\text S}(t)]_{21}$ as
\be
[\rho^{\text S}(t)]_{21}=[\rho_{I}^{\text S}(t)]_{21}e^{-2i\omega t}L_{12}(t),
\ee
where 
\be\label{eq-def-le}
L_{12}(t)=\frac{1}{\langle\psi_{I}^{1}(t)|\psi_{I}^{1}(t)\rangle}\langle\psi_{I}^{1}(t)|e^{iH_{\text B}^{(1)}t}e^{-iH_{\text B}^{(2)}t}|\psi_{I}^{1}(t)\rangle,
\ee
which can be regarded as a kind of Loschmidt echo(LE). 
We employ a further assumption which is
\begin{align}
L_{12}(t) \approx L^\text{typ}_{12}(t) & \equiv \langle \psi_\text{typ} |e^{iH_{\text B}^{(1)}t}e^{-iH_{\text B}^{(2)}t}|\psi_\text{typ}\rangle \nonumber \\
& \simeq \frac{1}{d}\text{Tr}_{\text B}\{e^{iH_{\text B}^{(1)}t}e^{-iH_{\text B}^{(2)}t}\}, \label{eq-asm-dec2}
\end{align}
where $|\psi_\text{typ}\rangle$ is a typical state in the Hilbert space of the bath.
 In this way, we arrive at the solution of equation of motion for the off-diagonal elements of the RDM in the Schr\"{o}dinger picture, which reads
\be\label{eq-rho12-le}
[\rho^{\text S}(t)]_{21}\simeq[\rho_{I}^{\text S}(t)]_{21}e^{-2i\omega t} L^\text{typ}_{12}(t).
\ee

In generic systems, the decay of LE is a complicated question, which has been investigated in a lot of works during the last decades\cite{LE-Casati02, LE-Cory02, LE-Haake01, LE-Jacquod01, LE-Jalabert02, LE-Peres84, LE-Prosen02, LE-Tomaz02, LE-Tomsovic02, LE-Wang05,gorin2006dynamics}.  	
However, some of the questions, especially for the LE decay in intermediate perturbation regime is still remain unclear. These are complicated questions and is not our intention to give exhaustive answers here.
At present we only focus on the weak ($\lambda_{mm}\ll 1$) and strong perturbation limit ($\lambda_{mm}\gg 1$).  
In the weak perturbation limit $\lambda_{mm}\ll 1$, it is found that, after a Gaussian decay at initial times, the LE follows a exponential decay (see \cite{gorin2006dynamics} and Refs. therein)
\be\label{eq-L12-weak}
|L^\text{typ}_{12}(t)| = \exp(-\Gamma_L t),
\ee
where
\begin{gather}\label{eq-GammaL}
\Gamma_{L}=\int_{0}^{\infty}\text{Tr}_{\text B}\{e^{iH_{\text B}\tau}\widetilde{V}e^{-iH_{\text B}\tau}\widetilde{V}\}{\text d}\tau,
\end{gather}
with
\be
\widetilde{V}=\lambda_{11}H_{11}-\lambda_{22}H_{22}.
\ee

In case of very strong perturbation ($\lambda_{mm}\gg 1$), the LE can be written as
\begin{align}
L_{12}^{\text{typ}}(t)\simeq & \langle\psi^{\text{typ}}|e^{i\lambda_{11}H_{11}t}e^{-i\lambda_{22}H_{22}t}|\psi^{\text{typ}}\rangle \nonumber \\
\simeq & \frac{1}{d}\text{Tr}_{\text B}\{e^{i\lambda_{11}H_{11}t}e^{-i\lambda_{22}H_{22}t}\}. \label{eq-L12-strong}
\end{align}
Making use of the eigenstates and eigenvalue of $H_{mm}$, denoted by $|E_k^{mm}\rangle$ and $E^{mm}_k$, respectively,  $L_{12}^\text{typ}(t)$ is written as
\be
L_{12}^{\text{typ}}(t)=\frac{1}{d}\sum_{k}e^{i\lambda_{11}E_{k}^{11}t}\langle E_{k}^{11}|e^{-i\lambda_{22}H_{22}t}|E_{k}^{11}\rangle.
\ee
As the interaction Hamiltonian $H_{11}$ and $H_{22}$ are uncorrelated, based on quantum typicality, one has
\be
\langle E_{k}^{11}|e^{-i\lambda_{22}H_{22}t}|E_{k}^{11}\rangle\simeq\frac{1}{d}\text{Tr}\{e^{-i\lambda_{22}H_{22}t}\},
\ee
which leads to 
\be\label{eq-L12-result0}
L_{12}^{\text{typ}}(t)\simeq\frac{1}{d^2}\text{Tr}\{e^{i\lambda_{11}H_{11}t} \}\text{Tr}\{ e^{-i\lambda_{22}H_{22}t} \}.
\ee
\subsection{Relation between relaxation and decoherence process}\label{sect-relation}
In open systems, for generic system-bath coupling, decoherence is induced by both \textit{EC} and \textit{EX} interactions.  The behavior of these two different kind of decoherence processes are usually different \cite{merkli2008dynamics}. 
So we will study their relation to the relaxation process separately.
In our case, the decoherence process induced by the \textit{EX} interaction is described by decoherence in the interaction picture, while the decoherence process induced by the \textit{EC} interaction is described by the Loschmidt echo introduced in Eq. \eqref{eq-def-le}.  
Decoherence in the Schr\"odinger picture is a joint effect of these two processes, which is described by Eq. \eqref{eq-rho12-le}.

The relation between decoherence induced by the \textit{EX} interaction and relaxation is quite simple, which can be seen from Eq. \eqref{eq-result-rdm}: the decoherence rate induced by the \textit{EX} interaction is half of the relaxation rate. 
However, the relation between decoherence induced by the \textit{EC} interaction and relaxation is not easily seen. Here we only consider the case of very strong \textit{EC} interaction $\lambda_{mm}\gg 1$.
In this case, the bath correlation function $F_{12}(\tau)$ can be written as 
\be
F_{12}(\tau)=\frac{1}{d}\text{Tr}_{\text B}\{e^{iH_{11}\tau}H_{12}e^{-iH_{22}\tau}H_{12}^{\dagger}\},
\ee
which in eigenstates of $H_{mm}$ reads
\be
F_{12}(\tau)=\frac{1}{d}\sum_{kl}e^{i(E_{k}^{11}-E_{l}^{22})\tau}|\langle E_{k}^{11}|H_{12}|E_{l}^{22}\rangle|^{2}.
\ee
If $H_{12}$ is traceless and uncorrelated with $H_{11}$ and $H_{22}$,  $\langle E_{k}^{11}|H_{12}|E_{l}^{22}\rangle$ can be regarded as Gaussian random numbers with mean zero. Then one can
replace $|\langle E_{k}^{11}|H_{12}|E_{l}^{22}\rangle|^2$ by its variance denoted by $\sigma_{12}$, which can be estimated in the following way
\be
\text{Tr}_{\text B}(H_{12}H_{12}^{\dagger})=\sum_{kl}|\langle E_{k}^{11}|H_{12}|E_{l}^{22}\rangle|^{2}\simeq d^{2}\sigma_{12},
\ee
yielding 
\be
\sigma_{12}\simeq\frac{1}{d^{2}}\text{Tr}_{\text B}\{H_{12}H_{12}^{\dagger}\}.
\ee
As a result, one has
\be
F_{12}(\tau)=\frac{1}{d^{2}}\text{Tr}_{\text B}\{H_{12}H_{12}^{\dagger}\}\frac{1}{d}\text{Tr}\{e^{i\lambda_{11}H_{11}\tau}\}\text{Tr}\{e^{-i\lambda_{22}H_{22}\tau}\}.
\ee
Comparing with Eq. \eqref{eq-L12-result0}, one can see that, the bath correlation function $F_{12}(t)$ is totally determined by the decoherence process induced by the \textit{EC} interaction, as
\be\label{eq-F12-00}
F_{12}(\tau)={\cal K}L_{12}^{\text{typ}}(\tau),
\ee
where 
\be
{\cal K} = \frac{1}{d}\text{Tr}_{\text B}\{H_{12}H_{12}^{\dagger}\},
\ee
which is the second moment of $H_{12}$.
Straightforwardly, one gets
\be\label{eq-relation}
\Gamma_{r}= 4{\cal K}|\lambda_{12}|^{2}\int_{0}^{+\infty}\mathrm{Re} [L_{12}^{\text{typ}}(\tau)]{\text d}\tau.
\ee
Under a corresponding ``uncorrelatedness assumption" (i.e. that $H_{mn}$ are uncorrelated)
the result of Eq. \eqref{eq-F12-00} can also be generated to a $N-$level system (see Appendix \ref{sect-mlevel} for more details).
Generally speaking a larger decay rate of $L^\text{typ}_{12}(\tau)$ results in a smaller time integral which in turn yields a smaller relaxation rate $\Gamma_r$. So it can be expected that the relaxation process will be slowed down by decoherence induced by the \textit{EC} interaction in case of $\lambda_{mm}\gg 1$. At the same time, decoherence induced by the \textit{EX} interaction is also suppressed by decoherence induced by the \textit{EC} interaction.

Before ending the section, it should be mentioned here that the simple relation between the relaxation and decoherence process induced by the \textit{EC} interaction (at large $\lambda_{mm}$) shown in Eq. \eqref{eq-relation} only exists under the assumption of the  interaction Hamiltonians to be uncorrelated, which can hardly be the case in realistic systems. In realistic systems, we expect certain relation between those two processes still exists, but it may take a more complicated form.




\section{Results in the spin random-matrix model}\label{sect-GORM}

In this section we are going to apply the results obtained in Sect. \ref{sect-RF} to a spin random-matrix model. The Hamiltonian is written as
\be\label{model-RMT}
H=\omega\sigma_{z}^{\text S}+H_{\text B}+\lambda_{d}V_{d}+\lambda_{r}V_{r},
\ee
where
\begin{align}
V_{d}= & |1\rangle\langle1|\otimes H_{11}+|2\rangle\langle2|\otimes H_{22}, \nonumber \\
V_{r}= & |1\rangle\langle2|\otimes H_{12}+|2\rangle\langle1|\otimes H_{12}^{\dagger},
\end{align}
and $H_{mn}$ and $H_{\text B}$ are modeled by random matrices, the elements of which are drawn from Gaussian distribution with zero mean and variance $\sigma_0^2 = \frac{1}{4d}$. Due to the Hermiticity of the total Hamiltonian,  $H_{\text B}$ and $H_{mm}$ should also be Hermitian, which means that they are actually drawn from Gaussian Orthogonal Ensemble(GOE). Moreoever, $\lambda_d$ and $\lambda_r$ should be real.
One can see that the Hamiltonian we consider here is a special case of the Hamiltonian defined in Eq. \eqref{eq-HSimple},  where $\lambda_{11} = \lambda_{22} = \lambda_d$, $\lambda_{12}=\lambda^*_{12}=\lambda_r$, and  \textit{EC} and \textit{EX} interactions are given by
\be
V_{\text{ec}} = \lambda_d V_d,\quad V_{\text{ex}} = \lambda_r V_r.
\ee
It should be mentioned here that this model is also very similar to the spin-GORM model introduced in Ref. \cite{esposito2003quantum}, except that $H_{12}$ considered here is not Hermitian.

\subsection{Time evolution of diagonal elements of RDM}
As shown in Eq. \eqref{eq-Gamma2}, the relaxation rate $\Gamma_r$ is determined by the bath correlation function
\be\label{eq-Fmn-def}
F_{12}(\tau)=\frac{1}{d}\text{Tr}_{\text B}\left\{ e^{i(-\omega+H_{\text B}^{(1)})\tau}H_{12}e^{-i(\omega+H_{\text B}^{(2)})\tau}H_{12}^{\dagger}\right\} .
\ee
Expanding $F_{12}(\tau)$ in the eigenbasis of $H_{\text B}^{(m)}$ yields
\be
F_{12}(\tau)=\frac{1}{d}\sum_{kl}e^{i(E_{k}^{(1)}-E_{l}^{(2)})\tau}e^{-2i\omega\tau}|\langle E_{k}^{(1)}|H_{12}|E_{l}^{(2)}\rangle|^{2}.
\ee
As $H_{12}$ is uncorrelated with $H_{\text B}$ and $H_{mm}$, the $\langle E_{k}^{(1)}|H_{12}|E_{l}^{(2)}\rangle$ can also be regarded as Gaussian random numbers with variance $\frac{1}{4d}$. As a result,
\begin{align}
F_{12}(\tau)  \simeq & \frac{1}{4d^{2}}e^{-2i\omega\tau}\text{Tr}_{\text B}\{e^{iH_{\text B}^{(1)}\tau}\}\text{Tr}_{\text B}\{e^{-iH_{\text B}^{(2)}\tau}\}\nonumber \\
 \simeq&\frac{1}{4d^{2}}e^{-2i\omega\tau}\int {\text d}E_{1}\Omega_{1}(E_{1})e^{-iE_{1}\tau} \nonumber \\
&\cdot \int {\text d}E_{2}\Omega_{2}(E_{2})e^{-iE_{2}\tau}\label{eq-F12-1},
\end{align}
where, as has already been defined, $\Omega_m(E)$ indicates the density of states of $H^{(m)}_{\text B}$.
Recalling the definition of $H^{(m)}_{\text B}$ in Eq. \eqref{eq-HBM}, one has
\be
H_{\text B}^{(1)}=H_{\text B}+\lambda_{d}H_{11},\ H_{\text B}^{(2)}=H_{\text B}+\lambda_{d}H_{22}.
\ee
As $H_{\text B}$ and $H_{mm}$ are uncorrelated, $H^{(1)}_{\text B}$ and $H^{(2)}_{\text B}$ can both be regarded as GOE random matrices, the  elements of which are drawn from the Gaussian distribution with mean zero and variance $\sigma_1=\sqrt{1+\lambda_d^2}\sigma_0$.
Thus, one has
\be
\Omega_{1}(E)\simeq\Omega_{2}(E)\simeq\Omega_{0}(E),
\ee
where $\Omega_0(E)$ has a semi-circle distribution \cite{haake2018quantum}
\begin{equation}\label{eq-DOS1}
\Omega_0(E)=\begin{cases}
\frac{2d}{\pi\sqrt{1+\lambda_{d}^{2}}}\sqrt{1-\frac{E^{2}}{1+\lambda_{d}^{2}}} & \text{for }|E|<\sqrt{1+\lambda_{d}^{2}}\\
0 & \text{otherwise}
\end{cases}.
\end{equation}
Substituting Eq. \eqref{eq-DOS1} into Eq. \eqref{eq-F12-1} and carrying out the integral, the bath correlation function can be written as
\be\label{eq-F12S}
F_{12}(\tau)=\frac{[{\cal J}_{1}(\alpha\tau)]^{2}}{(\alpha\tau)^{2}}e^{-2i\omega\tau},
\ee
where $\alpha = \sqrt{1+\lambda_d^2}$ and ${\cal J}_{1}(\tau)$ indicates the first order Bessel function of first kind. In case of $\omega \ll \alpha$ (which means the energy scale of the system is much smaller compared to that of the effective bath $H^{(m)}_{\text B}$), one can employ the rotating-wave-approximation (RWA). The phase factor $e^{-2i\omega \tau}$ can be approximated by $1$ (for time $\tau\lesssim\tau_{\text B}$) and $F_{12}(\tau)$ is only dependent on the rescaled time $\alpha \tau$ as
\be\label{eq-F12-final}
F_{12}(\tau) = \frac{[{\cal J}_{1}(\alpha\tau)]^{2}}{(\alpha\tau)^{2}}.
\ee
Inserting Eq. \eqref{eq-F12-final} to Eq. \eqref{eq-Gamma2} and carrying out the integral, one has
\be\label{eq-result-I12}
\int_{-\infty}^{\infty}F_{12}(\tau)\text{d}\tau  = \frac{8}{3\alpha\pi}=\frac{8}{3\sqrt{1+\lambda_{d}^{2}}\pi},
\ee
which leads to
\be\label{eq-result}
\Gamma_r \approx \frac{16\lambda^2_r}{3\pi\sqrt{1+\lambda^2_d}} \propto\frac{\lambda^2_r}{\sqrt{1+\lambda^2_{d}}}.
\ee
Thus one has

\be\label{eq-result-dg}
\rho_{11}^{\text S}(t)-\rho_{11}^{{\text S}}(\text{eq})=e^{-\frac{16\lambda_{r}^{2}}{3\pi\sqrt{1+\lambda_{d}^{2}}}t}(\rho_{11}^{\text S}(0)-\rho_{11}^{{\text S}}(\text{eq})).
\ee
It implies that relaxation is boosted by the \textit{EX} interaction and suppressed by the \textit{EC} interaction.

It should be mentioned here that, with the traditional approach where only the non-interacting part ($H_{\text S}+H_{\text B}$) of the Hamiltonian is treated non-perturbatively, a similar exponential decay $\rho_{11}^{\text S}(t)-\rho_{11}^{\text S}(\text{eq}) \propto e^{-\Gamma^0_r t}$ can be derived. The different is that, in that case one has $\Gamma^0_r=-\frac{16\lambda_{r}^{2}}{3\pi}$, which is independent of the \textit{EC} interaction strength $\lambda_d$. This indicates that the traditional approach is unable to account for impact of decoherence on the relaxation process. It is not surprising, as the traditional approach is only supposed to work in the weak-coupling regime where not only the \textit{EX} interaction but also the \textit{EC} interaction should be weak. In this regime ($\lambda_d \ll 1, \lambda_r \ll 1$), one can easily see that $\Gamma_r \approx \Gamma^0_r$ and the results of the two approaches agree with each other.

After having derived the relaxation rate, we come back
to Eq. \eqref{eq-condition-0}  to check under what condition it is fulfilled. 
Here, for example we can define $\tau_B$ and $\tau_R$ to be the time at which $\mathrm{Re}[F_{12}(\tau)]$ and $|\rho_{11}^{\text S}(t)-\rho_{11}^{\text S}(\text{eq})|$ decay to $1\%$ of its initial value and never exceed that value afterwards. Then with straightforward calculations one has
\be\label{eq-tauFR}
\tau_B \approx \frac{5.9}{\sqrt{1+\lambda_{d}^{2}}} ,\quad \tau_R = \frac{2\ln10}{\Gamma_{r}}\approx\frac{2.7\sqrt{1+\lambda_{d}^{2}}}{\lambda_{r}^{2}}.
\ee
As a result,  Eq. \eqref{eq-condition-0} can be approximately rewritten as
\be
\lambda_{r}\ll\sqrt{\frac{1+\lambda_{d}^{2}}{2}},
\ee
indicating that the Eq. \eqref{eq-condition-0} would be better fulfilled for larger $\lambda_d$, if $\lambda_r$ is fixed.

\subsection{Time evolution of off-diagonal elements of RDM}
As $\Gamma_r$ has already been derived in Eq. \eqref{eq-result}, the time evolution of off-diagonal elements of RDM in the interaction picture can be written as
\be\label{eq-result-nd}
|[\rho_{I}^{\text S}(t)]_{21}|=e^{-\frac{8\lambda^2_{r}}{3\pi\sqrt{1+\lambda_{d}^{2}}}t}|[\rho_{I}^{\text S}(0)]_{21}|,
\ee
from which one can see that decoherence in the interaction picture (or decoherence induced by the \textit{EX} interaction) is also suppressed by the \textit{EC} interaction.

In the Schr\"{o}dinger picture, the time evolution of off-diagonal elements of RDM is given in Eq. \eqref{eq-rho12-le}. As $|[\rho_{I}^{\text S}(t)]_{21}|$ has already been given above, one only needs to study the LE term $L_{12}^\text{typ}(t)$ which characterizes the  decoherence process induced by the \textit{EC} interaction.  	
In the weak perturbation limit $\lambda_{mm}\ll 1$, as has already been discussed in Sec. \ref{sect-RDMS}, after a Gaussian decay at initial times, the LE decays exponentially
\be
|L^\text{typ}_{12}(t)| = \exp(-\Gamma_L t),
\ee
where
\be\label{eq-GammaL}
\Gamma_{L}=\lambda_{d}^{2}\int_{0}^{\infty}{\text d}\tau F_{dd}(\tau),
\ee
and
\begin{gather}
F_{dd}(\tau)\equiv\frac{1}{d}\text{Tr}_{\text B}\{e^{iH_{\text B}\tau}H_{d}e^{-iH_{\text B}\tau}H_d\} \\
H_d \equiv H_{11} - H_{22}.
\end{gather}
Expanding $H_d$ in the eigenbasis of $H_{\text B}$, one has
\be
F_{dd}(\tau)=\frac{1}{d}\sum_{kl}e^{i(E_{k}-E_{l})\tau}|\langle E_{k}|H_{d}|E_{l}\rangle|^{2}.
\ee
As $H_d$ is also a (GOE) random matrix,  $|\langle E_{k}|H_{d}|E_{l}\rangle|^{2}$ can be replaced by its variance $\frac{1}{2d}$, which yields
\be
F_{dd}(\tau) = \frac{1}{2d^2}|\text{Tr}_B\{e^{-iH_{\text B} \tau}\}|^2 = \frac{2[{\cal J}_{1}(\tau)]^{2}}{\tau^{2}}.
\ee
Carrying out the integral in Eq. \eqref{eq-GammaL} exactly, one gets
\be
\Gamma_L = \frac{8\lambda_{d}^{2}}{3\pi}.
\ee

\begin{figure}[tb]
		\includegraphics[width=0.95\columnwidth]{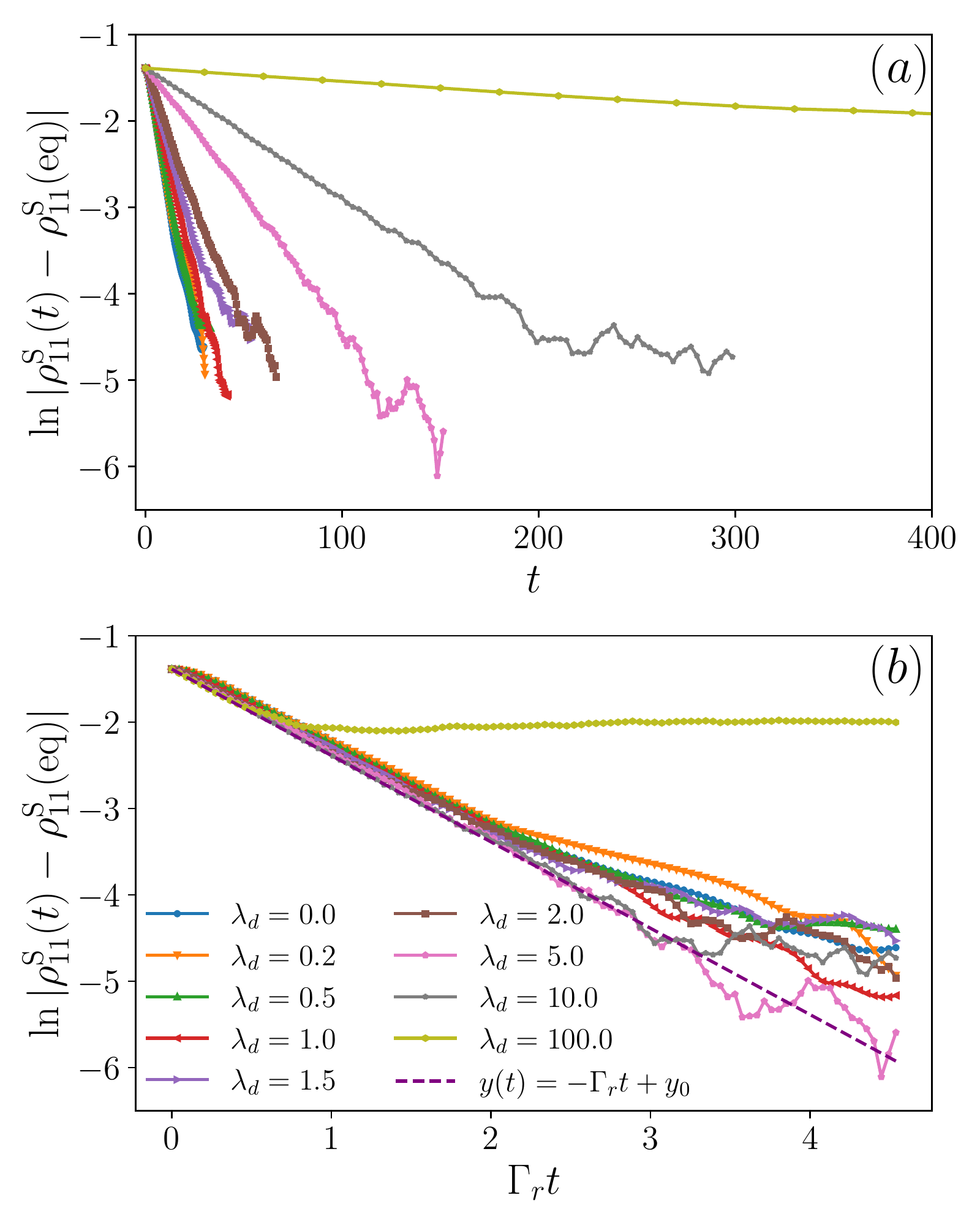}
	\caption{(a) $\ln|\rho_{11}^{\text S}(t)-\rho_{11}^{\text S}(\text{eq})|$ as a function of (a) $ t$ and (b) $\Gamma_r t$, for different $\lambda_d$, where $\Gamma_r$ is given in Eq. (\ref{eq-result}), for  $\lambda_{r}=0.3, \omega = 0.05, d=2^{14}$. The dashed line indicates the analytical prediction $\ln|\rho_{11}^{\text S}(t)-\rho_{11}^{\text S}(\text{eq})| = -\Gamma_r t + y_0$, where $y_0=\ln|\rho_{11}^{\text S}(0)-\rho_{11}^{\text S}(\text{eq})|$.}
		\label{Fig-Diag-D}
	\end{figure}

In case of $\lambda_d \gg 1$, one has
\begin{align}
L_{12}^{\text{typ}}(t)\simeq & \frac{1}{d}\text{Tr}_{\text B}\{e^{i\lambda_{d}H_{11}t}e^{-i\lambda_{d}H_{22}t}\}\nonumber \\
\simeq & \frac{1}{d^2}\text{Tr}_{\text B}\{e^{i\lambda_{d}H_{11}t}\}\text{Tr}_{\text B}\{e^{-i\lambda_{d}H_{22}t}\}\nonumber \\
\simeq & \frac{1}{d^2}|\text{Tr}_{\text B}\{e^{i\lambda_{d}H_{11}t}\}|^2 \simeq \frac{4[{\cal J}_{1}(\lambda_{d}t)]^{2}}{(\lambda_{d}t)^{2}}, \label{eq-L12-strong}
\end{align}
where the second line is obtained due to the fact that $H_{11}$ and $H_{22}$ are uncorrelated.
As we have discussed in Sec. \ref{sect-relation}, by comparing Eq. \eqref{eq-L12-strong} with Eq. \eqref{eq-F12-final}, one finds
\be\label{eq-relation-FL}
F_{12}(t)=\frac{1}{4}L_{12}^{\text{typ}}(t),\text{ for }\lambda_{d}\gg1,
\ee
indicating that the bath correlation function $F_{12}(t)$ is determined by the decoherence process induced by the \textit{EC} interaction. We stress here again that  the simple relation between $F_{12}(t)$ and $L^\text{typ}_{12}(t)$ in Eq. \eqref{eq-relation-FL} for $\lambda_d \gg 1$ relies on the assumption that $H_{12}$, $H_{22}$ and $H_{11}$ are all uncorrelated. To see what would happen if they are not fully uncorrelated,  here as an example one can relax one of the restrictions, by considering the case $H_{22}=-H_{11}$,  while still assuming they are uncorrelated with $H_{12}$. With similar derivations, one can see that $F_{12}(t)\simeq\left(\frac{{\cal J}_{1}(\lambda_{d}t)}{\lambda_{d}t}\right)^{2}$ which remains the same, while $L^\text{typ}_{12}(t) \propto \frac{{\cal J}_{1}(2\lambda_{d}t)}{\lambda_{d}t}$, different from the fully uncorrelated case. In this case although $F_{12}(t)$ can not be directly written as a function of $L^\text{typ}_{12}(t)$, they are still strongly correlated. It is also to be expected that a faster decay of $L^\text{typ}_{12}(t)$ results in a smaller relaxation rate $\Gamma_r$.

In summary, one has
\be\label{eq-result-le}
|L_{12}^{\text{typ}}(t)|\simeq\begin{cases}
\exp(-\frac{8}{3\pi}\lambda_{d}^{2}t), & \lambda_{d}\ll1\\
\frac{4[{\cal J}_{1}(\lambda_{d}t)]^{2}}{(\lambda_{d}t)^{2}}. & \lambda_{d}\gg1
\end{cases}.
\ee
From the result in Eq. \eqref{eq-result-le}, one can see that in both cases, the decay of $L^\text{typ}_{12}(t)$, which characterize the decoherence process induced by the \textit{EC} interaction, becomes faster for larger $\lambda_d$.
Substituting Eq. \eqref{eq-result-le} into Eq. \eqref{eq-rho12-le}, one derives the equation of motion for off-diagonal elements of RDM in Schr\"{o}dinger picture
\be\label{eq-result-dec0}
|\rho_{12}^{\text S}(t)|\simeq\begin{cases}
\exp(-\frac{8}{3\pi}(\frac{\lambda_{r}^{2}}{\sqrt{1+\lambda_{d}^{2}}}+\lambda_{d}^{2})t)|\rho_{12}^{\text S}(0)|, & \lambda_{d}\ll1 \\
\exp(-\frac{8}{3\pi}\frac{\lambda_{r}^{2}}{\sqrt{1+\lambda_{d}^{2}}}t)\frac{4[{\cal J}_{1}(\lambda_{d}t)]^{2}}{(\lambda_{d}t)^{2}}|\rho_{12}^{\text S}(0)|. & \lambda_{d}\gg1
\end{cases}
\ee
In both $\lambda_d \ll 1$ and $\lambda_d \gg 1$ cases, it can be seen that, the decoherence time $\tau_D$ decreases with increasing \textit{EX} coupling strength $\lambda_r$.
For $\lambda_d \gg 1$, one can actually approximate $\exp(-\frac{8}{3\pi}\frac{\lambda_{r}^{2}}{\sqrt{1+\lambda_{d}^{2}}}t)$ by $1$, and in this case 
\be\label{eq-rho12-large}
|\rho_{21}^{\text S}(t)|\simeq\frac{4[{\cal J}_{1}(\lambda_{d}t)]^{2}}{(\lambda_{d}t)^{2}}|\rho_{21}^{\text S}(0)|,
\ee
which is a function of $\lambda_d t$. One can conclude that $\tau_D\propto \frac{1}{\lambda_d}$.  For $\lambda_d \ll 1$, $|\rho_{12}^{\text S}(t)|$ follows an exponential decay, with a decoherence rate $\Gamma_d$ given by 
\be\label{eq-GammaD}
\Gamma_{d}=-\frac{8}{3\pi}(\frac{\lambda_{r}^{2}}{\sqrt{1+\lambda_{d}^{2}}}+\lambda_{d}^{2}).
\ee
Expanding $\frac{1}{\sqrt{1+\lambda_{d}^{2}}}$ to the second order of $\lambda_d$,
one finds that 
$\Gamma_d \propto \lambda^2_d(1-\frac{\lambda^2_r}{2})$, which is a monotonically increasing function when $\lambda_r < \sqrt{2}$, and becomes a monotonically decreasing function of $\lambda_d$ when $\lambda_r > \sqrt{2}$.  Combining with the condition for the Markov approximation $\lambda_{r}\ll\sqrt{\frac{1+\lambda_{d}^{2}}{2}}$ in Eq. \eqref{eq-condition-0}, one expects in situations where the analytical result in Eq. \eqref{eq-result-dec0} is applicable, $\Gamma_d$ is a monotonically increasing function of $\lambda_d$. In summary, we find that in case of $\lambda_d \ll 1$ and $\lambda_d \gg 1$, decoherence in the Schr\"{o}dinger picture is boosted by both \textit{EX} and \textit{EC} interactions.

\begin{figure}[tb]
		\includegraphics[width=0.95\columnwidth]{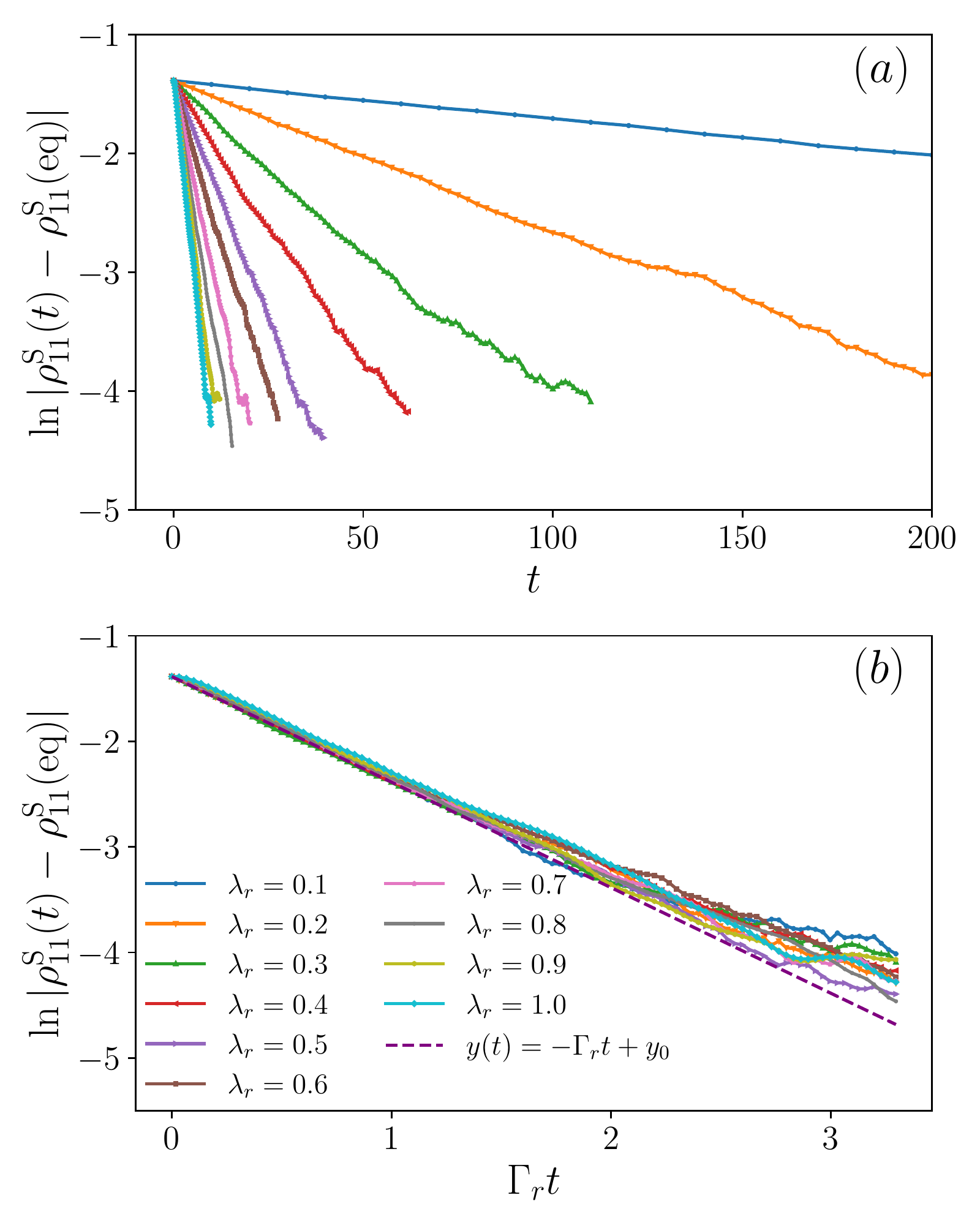}
	\caption{Similar to Fig. \ref{Fig-Diag-D}, but results for varying $\lambda_r$ with fixed a $\lambda_d$, where $\lambda_{d}=5.0, \omega = 0.05, d=2^{14}$.}
		\label{Fig-Diag-R}
	\end{figure}




\section{Numerical results}\label{sect-numerical}
\subsection{Main results on relaxation and decoherence dynamics}
In this section, we numerically checked our analytical results on the relaxation and decoherence processes in the spin Random-Matrix model, which are given in Eqs. \eqref{eq-result-dg} and \eqref{eq-result-dec0}.
In the numerical simulations, as in Eq. \eqref{eq-initial-pure}, we consider a pure state as initial state, which is written as
\be
|\psi(0)\rangle = |\psi^{\text S}(0)\rangle \otimes |\psi^{\text B}(0)\rangle.
\ee
Here $|\psi^{\text S}(0)\rangle=\frac{\sqrt{3}}{2}|1\rangle+\frac{1}{2}|2\rangle$, and $|\psi^{\text B}(0)\rangle$ is a random state in the bath Hilbert space

\be
|\psi^{\text B}(0)\rangle=\sum_{k=1}^{d}c_{k}|E_{k}\rangle.
\ee
$c_k$ are complex numbers, the real and imaginary parts of which are drawn independently from Gaussian distribution, and due to the normalization of $|\psi^B(0)\rangle$, $\sum_{k=1}^d |c_k|^2 = 1$.

The time evolution of diagonal elements of RDM are studied in Fig. \ref{Fig-Diag-D} and Fig.  \ref{Fig-Diag-R}, where the results are shown for varying $\lambda_d$ with a fixed $\lambda_r$ and varying $\lambda_r$ with a fixed $\lambda_d$, respectively. 
One can see that, in both figures, the resulting curves approximately overlap as functions of $\Gamma_rt$ ($\Gamma_r$ is given in Eq. \eqref{eq-result}), which also agrees with the analytical prediction in Eq. \eqref{eq-result-dg}, at least up to a certain time scale.
The relaxation rate is found to increase with $\lambda_r$ while it decreases with $\lambda_d$ (Fig. \ref{Fig-Diag-R}(a) and Fig. \ref{Fig-Diag-D}(a)). It indicates that the relaxation process is boosted by the \textit{EX} interaction, and  suppressed by the \textit{EC} interaction at the same time.

We notice that, for extremely strong $\lambda_d$, e.g., $\lambda_d = 100$, the numerical results only follow the analytical prediction for very short time and then goes to a non-thermal steady value. This is due to a finite-size effect. 
 In systems with finite Hilbert space dimension $d$, for sufficiently large $\lambda_d$, the off-diagonal elements of the \textit{EC} interaction $\lambda_r V_r$ in the eigenbasis of the unperturbed Hamiltonian $H_0 = H_{\text S} + H_{\text B} + \lambda_d V_d$ is not much larger, or even smaller than the average level spacing of $H_0$ . As a result, the system can not thermalize to the infinite temperature Gibbs state (which is just the unitary matrix), but to a non-thermal steady state which usually depends on the initial state. But as the off-diagonal elements of $\lambda_r V_r$ scale as $\frac{1}{\sqrt{d}}$ while the mean level spacing of $H_0$ scales as $\frac{1}{d}$, such finite-size effect will vanish if $d$ is large enough. Thus we expect that if we continue to increase $d$, the result for $\lambda_d = 100$ would follow the analytical prediction for longer time.

\begin{figure}[tb]
		\includegraphics[width=0.95\columnwidth]{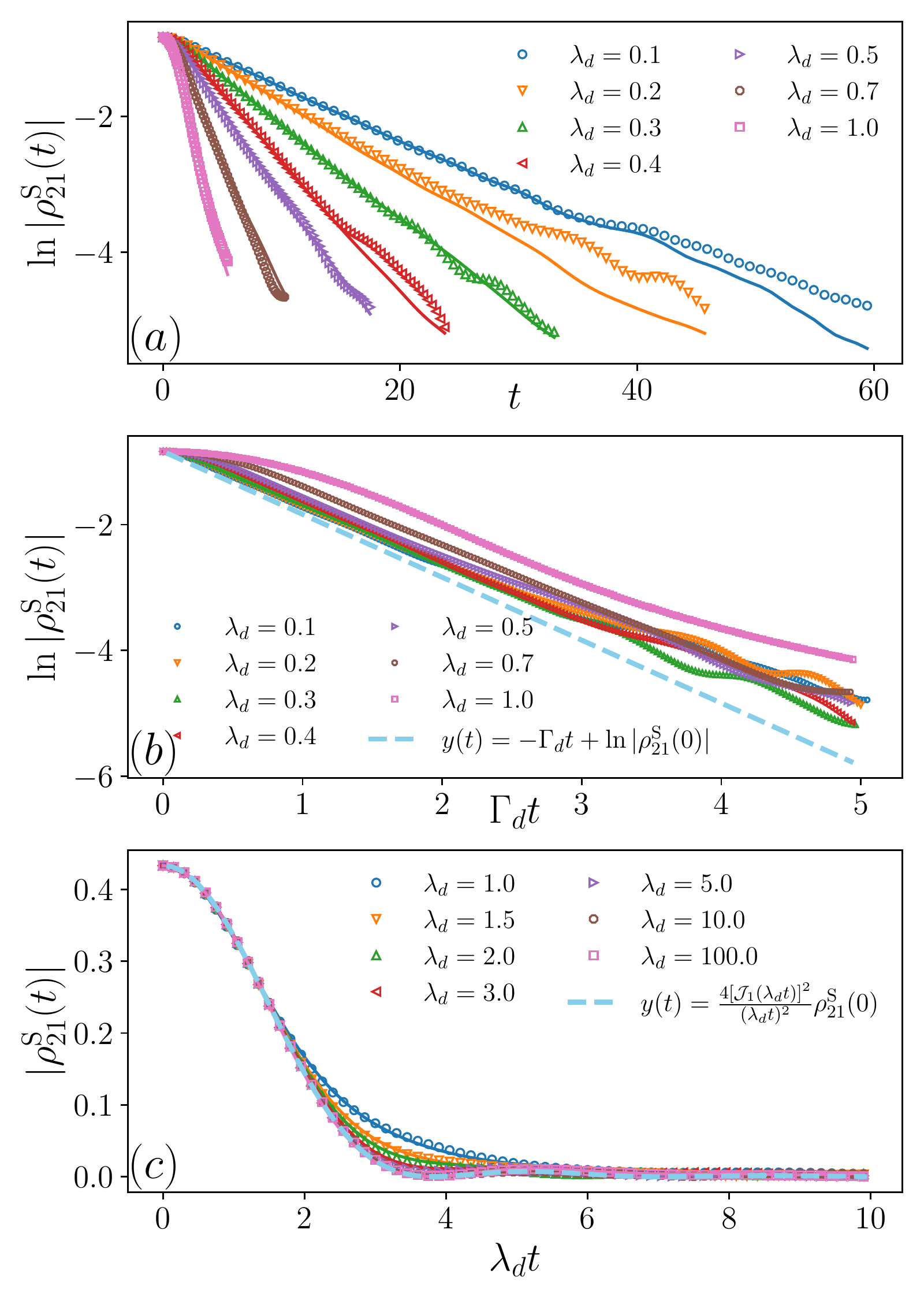}
	\caption{Decoherence in the Schr\"{o}dinger picture: (a) $\ln|\rho_{21}^{\text S}(t)|$ versus $ t$, (b) $\ln|\rho_{21}^{\text S}(t)|$ versus $ \Gamma_d t$ and (c) $|\rho_{21}^{\text S}(t)|$ versus $\lambda_d t$, for different $\lambda_d$.  The dashed line in (b) 
    shows the analytical prediction for $\lambda_d \ll 1$, $\ln|\rho_{21}^{\text S}(t)|=-\Gamma_{d}t+\ln|\rho_{21}^{\text S}(0)|$ ,where $\Gamma_d$ is given in Eq. \eqref{eq-GammaD}.
    Solid line in (a) and (c) indicates the semi-analytical prediction in Eq. \eqref{eq-rho12-le}, where both $[\rho^{\text S}_I(t)]_{21}$ and $L^\text{typ}_{12}(t)$ are calculated numerically as well (shown in Fig. \ref{Fig-ND-D} and Fig. \ref{Fig-LE}). The dashed line in (c) indicates the analytical prediction for $\lambda_d \gg 1$, $|\rho_{21}^{\text S}(t)|\simeq\frac{4[{\cal J}_{1}(\lambda_{d}t)]^{2}}{(\lambda_{d}t)^{2}}|\rho_{21}^{\text S}(0)|$.}
		\label{Fig-Dec-Small}
	\end{figure}


\begin{figure}[tb]
		\includegraphics[width=0.95\columnwidth]{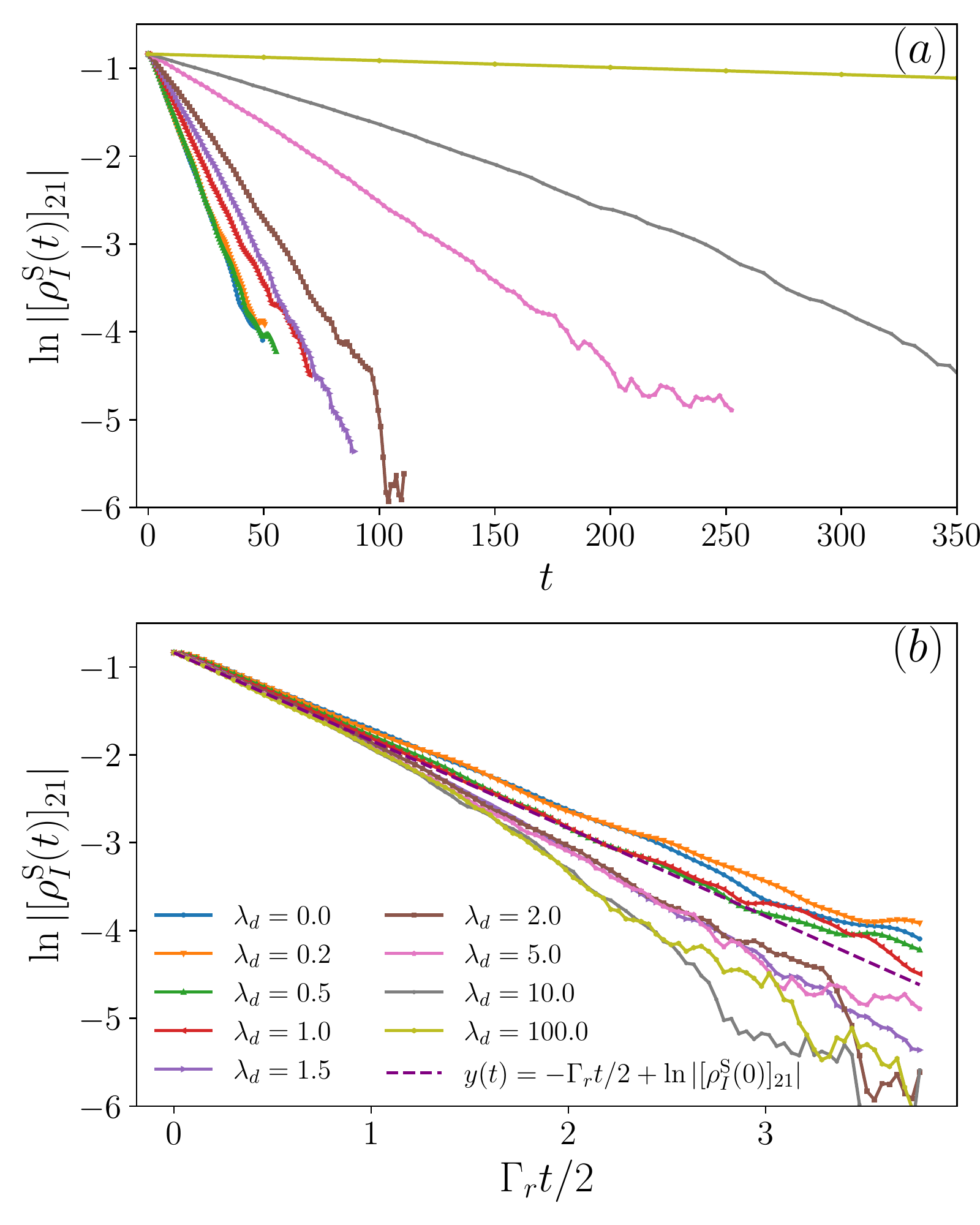}
	\caption{Decoherence induced by the \textit{EX} interaction (decoherence in the interaction picture): $\ln|[\rho_{I}^{\text S}(t)]_{21}|$ as a function of (a) $t$ and (b) $\Gamma_r t/2  $, for different $\lambda_d$, where $\Gamma_r$ is given in Eq. (\ref{eq-result}). The dashed line in (b) indicates the analytical prediction $\ln|[\rho_{I}^{\text S}(t)]_{21}|=-\Gamma_r t/2+\ln|[\rho_{I}^{\text S}(0)]_{21}|$.
 Here $\lambda_{r}=0.3, \omega = 0.05, d=2^{14}$.}
		\label{Fig-ND-D}
	\end{figure}

Results for decoherence process in the Schr\"{o}dinger picture are shown in Fig. \ref{Fig-Dec-Small}. One can see from Fig. \ref{Fig-Dec-Small}(a) and Fig. \ref{Fig-Dec-Small}(c) that, our semi-analytical predictions in Eq. \eqref{eq-rho12-le} works quite well in both cases. It implies that the two assumptions we made in Eq. \eqref{eq-asm-dec1} and Eq. \eqref{eq-asm-dec2} are reasonable.
Moreover, the analytical prediction for $\lambda_d \ll 1$ and $\lambda_d \gg 1$ in Eq. \eqref{eq-result-dec0} are found to agree quite well with the numerical results. Surprisingly, the analytical prediction for weak $\lambda_d$ is even valid for intermediate $\lambda_d \approx 1.0$.  For all values of $\lambda_d$ we consider here, one can see that decoherence become faster if one increases $\lambda_d$, indicating that decoherence process is always boosted by the \textit{EC} interaction, which is just what one would expect.


\begin{figure}[tb]
		\includegraphics[width=0.95\columnwidth]{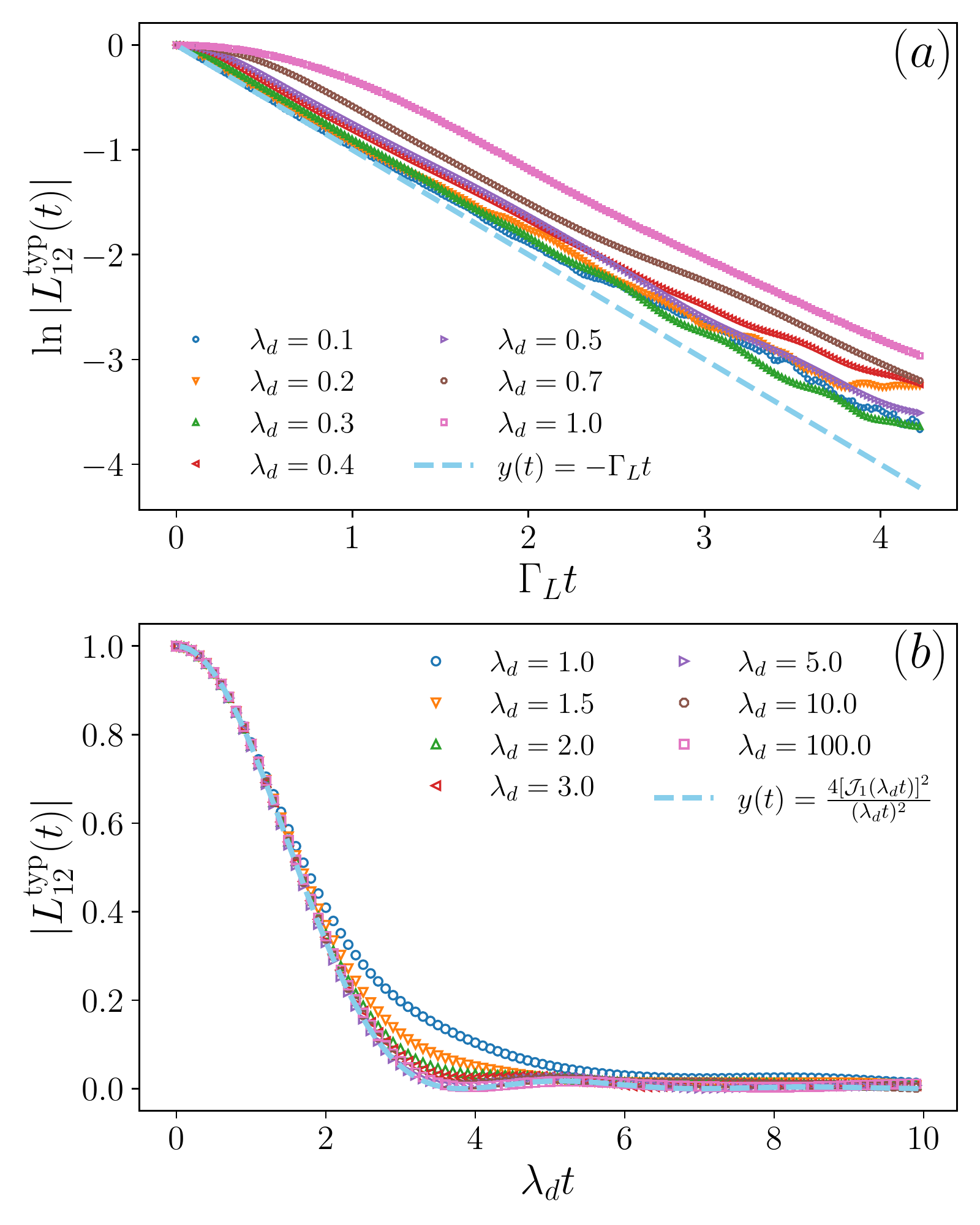}
	\caption{Decoherence induced by the \textit{EC} interaction: (a) $\ln|L^\text{typ}_{12}(t)|$ versus $\Gamma_L t$, (b) $|L^\text{typ}_{12}(t)|$ versus $\lambda_d t$, for different $\lambda_d$. The solid line in (a) 
    shows the analytical prediction for $\lambda_d \ll 1$, $\ln(|L^\text{typ}_{12}(t)|) = -\Gamma_L t$, where $\Gamma_{L}=\frac{8}{3\pi}{\lambda_{d}^{2}}$.
    The dashed line in (b) indicates the analytical prediction for $\lambda_d \gg 1$, $|L^\text{typ}_{12}(t)|=\frac{4[{\cal J}_{1}(\lambda_{d}t)]^{2}}{(\lambda_{d}t)^{2}}$.   Here $\lambda_{r}=0.3, \omega = 0.05, d=2^{14}$.}
		\label{Fig-LE}
	\end{figure}

\subsection{Additional results on the bath correlation function and decoherence induced by \textit{EX} and \textit{EC} interactions}
Additionally, we study the decoherence process induced by \textit{EX} and \textit{EC} interactions separately, the results are shown in  Fig. \ref{Fig-ND-D} and Fig. \ref{Fig-LE} .
Results for decoherence induced by the \textit{EX} interaction (which is described by decoherence in the interaction picture) are shown in Fig.  \ref{Fig-ND-D}.
A good agreement with  the analytical prediction $\ln|[\rho^{\text S}_I(t)]_{21}|=-\frac{\Gamma_r}{2}t$ up to a certain time scale can be seen. Similar to the relaxation process, one finds in Fig. \ref{Fig-ND-D}(a) that, the decoherence process induced by the \textit{EX} interaction is also suppressed by the \textit{EC} interaction.

In Fig. \ref{Fig-LE}, we show results for the decoherence process induced by the \textit{EC} interaction, which is described by $L^\text{typ}_{12}(t)$, for a wide range of $\lambda_d$. A good agreement with the analytical prediction in Eq. \eqref{eq-result-le}  can be found, for both $\lambda_d \ll 1$ and $\lambda_d \gg 1$. Based on the results shown in Fig. \ref{Fig-LE}, one finds that $L^\text{typ}_{12}(t)$  decays faster for larger $\lambda_d$, indicating that the decoherence process induced by the \textit{EC} interaction is always boosted by the \textit{EC} interaction, just as one expects.

\begin{figure}[t!]
		\includegraphics[width=0.95\columnwidth]{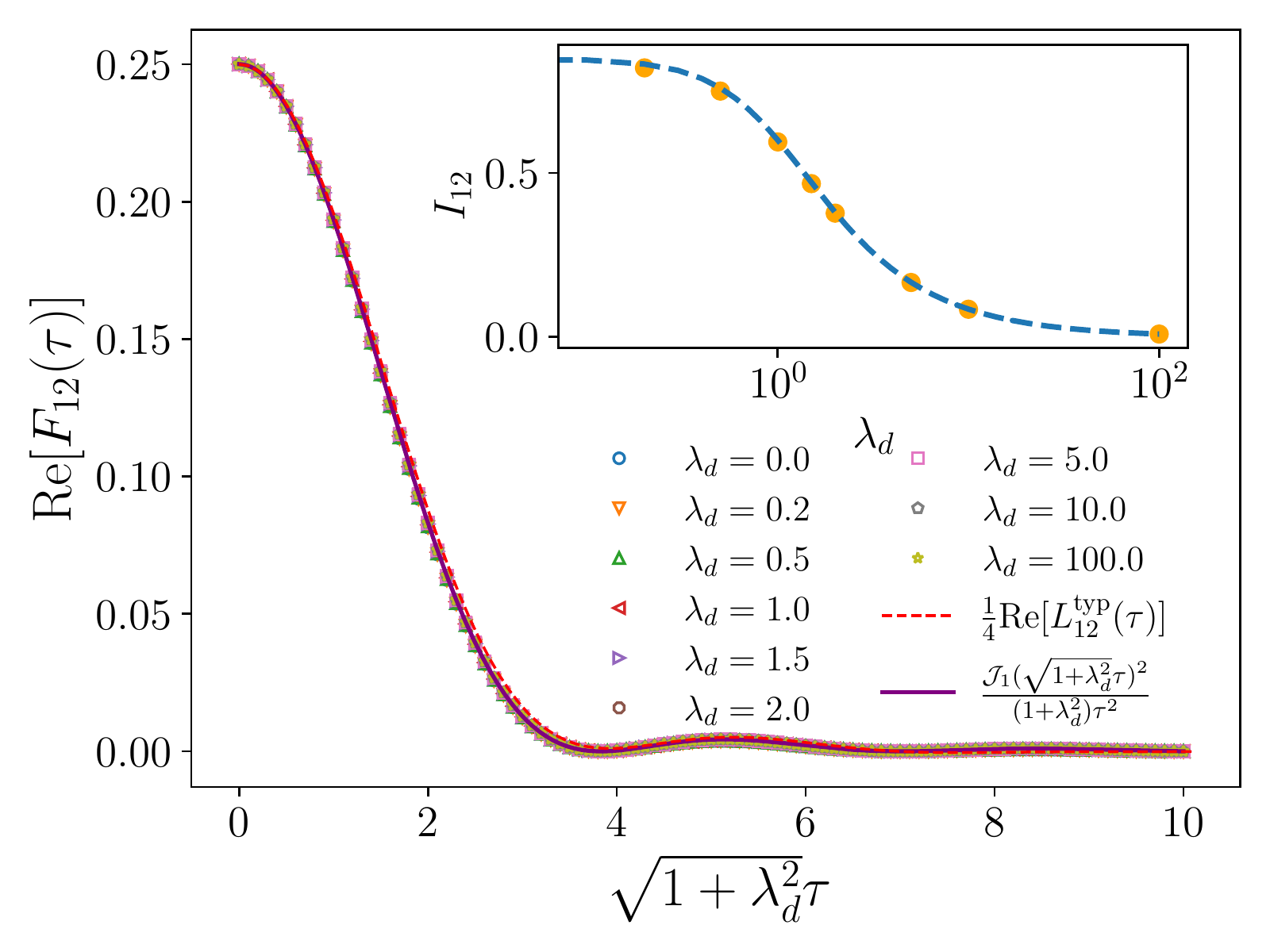}
	\caption{The real part of correlation function $F_{12}(\tau)$ versus rescaled time $\sqrt{1+\lambda^2_d}\tau$ for different $\lambda_d$ for $\omega = 0.05$ and $d = 2^{14}$. The solid line indicates the analytical prediction in Eq. \eqref{eq-F12S}, and the dashed line shows the real part of the Loschmidt echo $\frac{1}{4}L^\text{typ}_{12}(t)$ for $\lambda_d = 5$.
    The inset shows the infinite time integral $ I_{12}=\int_{-\infty}^{\infty}F_{12}(\tau)d\tau$ where the dashed line indicates the analytical prediction in Eq. \eqref{eq-result-I12}}
		\label{Fig-BBt}
	\end{figure}

The bath correlation function $F_{12}(\tau)$ is also calculated, and the results are shown in Fig. \ref{Fig-BBt}. $F_{12}(\tau)$ for different $\lambda_d$ overlap as functions of $\sqrt{1+\lambda^2_d}$, which agrees with the analytical prediction in Eq. \eqref{eq-F12S} as well. A good agreement between $F_{12}(\tau)$ and $\frac{1}{4}L^\text{typ}_{12}(\tau)$ (for $\lambda_d = 5$) can also be seen, which confirms the analytical result in Eq. \eqref{eq-relation-FL}. It indicates that, in the spin Random-Matrix model we consider here, the correlation function $F_{12}(\tau)$ is  determined by the decoherence process induced by the \textit{EC} interaction in case of $\lambda_d \gg 1$.
In the inset, the numerical results of infinite time intergral of $F_{12}(\tau)$ fits perfectly with the analytical prediction in Eq. \eqref{eq-result-I12}, from which one concludes that the relaxation rate $\Gamma_r$ becomes smaller for larger $\lambda_d$. 

In summary, our main results for the reduced equations of motion for the qubit in the spin Random-Matrix model are confirmed in a wide range of coupling strength $\lambda_r$ and $\lambda_d$. 
A simple relation between the bath correlation function and the decoherence process induced by the \textit{EC} interaction can be seen for $\lambda_d \gg 1$.
Moreover, we find that, if one increases $\lambda_d$,  decoherence induced by the \text{EC} interaction becomes faster
while relaxation as well as decoherence  induced by the \textit{EX} interaction become slower.

\section{Conclusions and Outlook}\label{Sect-Con}
In this paper, by employing the modified Redfield theory, we derived reduced equations of motion for the qubit in a spin random-matrix model, which is also valid for large \textit{EC} interactions. The relation between the relaxation and decoherence process is discussed. 
We find a simple relation between the bath correlation function and the decoherence process induced by the \textit{EC} interaction in strong decoherence regime.
It implies that the relaxation process is suppressed by decoherence  induced by the \textit{EC} interaction. The relaxation rate goes to zero, if the \textit{EC} interaction is sufficiently strong, which coincides with the quantum zeno effect.  Furthermore,  decoherence  induced by the \textit{EX} interaction is also found to be suppressed by decoherence induced by the $\textit{EC}$ interaction. 

As our main results are derived in a idealistic model where all the interaction Hamiltonians are assumed to be uncorrelated, it is interesting to ask whether or to what extent our finding could be applied to realistic systems, which will be investigated in our future work.
 
\section{Acknowledgement}

JW thanks Wen-ge Wang and Hua Yan for interesting discussion on this topic.
This work has been funded by the Deutsche Forschungsgemeinschaft (DFG), under
Grant No.\ 397107022, No.\ 397067869, and No.\ 397082825,
within the DFG Research Unit FOR 2692, under Grant No.\ 355031190.

\bibliographystyle{apsrev4-1}
\bibliography{Ref}

\begin{appendix}

\section{Derivation of reduced motion of RDM in the interaction picture}\label{sect-derivation}

In this section we show detailed derivations of the reduced equations of motion for the system in the interaction picture, where
we start from Eq. \eqref{eq-drho0} in the main text,
\be
\frac{{\text d}}{{\text d}t}\rho_{I}^{\text S}(t)=-\frac{1}{d}\int_{0}^{t}{\text d}s\text{Tr}_{\text B}\left\{ [H^{\text{int}}_{I}(t),[H^{\text{int}}_{I}(s),\rho_{I}^{\text S}(s)]]\right\}.
\ee

For the convenience of the discussions below, we divide $\frac{d}{dt}\rho^{\text S}(t)$ into four terms as,
\be\label{eq-drhoS}
\frac{{\text d}}{{\text d}t}\rho_I^{\text S}(t)=S_1 + S_2 + S^\dagger_1 + S^\dagger_2,
\ee
where 
\begin{align}
S_1 & = -\frac{1}{d}\int_{0}^{t}{\text d}s\text{Tr}_{\text B}\left\{ H^{\text{int}}_{I}(t)H^{\text{int}}_{I}(s)\rho^{\text S}_{I}(s)\right\},  \\
S_2 & = \frac{1}{d}
\int_{0}^{t}{\text d}s\text{Tr}_{\text B}\left\{ H^{\text{int}}_{I}(t)\rho^{\text S}_{I}(s)H^{\text{int}}_{I}(s)\right\}.
\end{align}
Here and in the rest of the section, we omit the subscript $I$ for simplicity.

First we start from $S_1$, which can be written as 
\be\label{eq-drhoA}
S_{1}  =-\frac{1}{d}\int_{0}^{t}{\text d}s\text{Tr}_{\text B}\left\{ H^{\text{int}}(t)H^{\text{int}}(s)\right\} \rho^{\text S}(s).
\ee
Inserting Eq.(\ref{eq-VI}), one has
\begin{gather}
S_1   = \nonumber \\
-\frac{1}{d}|\lambda_{12}|^2|1\rangle\langle1|\int_{0}^{t}{\text d}s\text{Tr}_{\text B}\left\{ U^\dagger_{1}(t-s)H_{12}U_{2}(t-s)H_{12}^{\dagger}\right\}\rho^{\text S}(s) \nonumber \\
 -\frac{1}{d}|\lambda_{12}|^2||2\rangle\langle2|\int_{0}^{t}{\text d}s\text{Tr}_{\text B}\left\{ U^\dagger_{2}(t-s)H_{12}^\dagger U_{1}(t-s)H_{12}\right\}\rho^{\text S}(s) \nonumber \\
=   -|\lambda_{12}|^2|1\rangle\langle1|\int_{0}^{t}F_{12}(t-s)\rho^{\text S}(s){\text d}s \nonumber  \\ -|\lambda_{12}|^2|2\rangle\langle2|\int_{0}^{t}F_{21}(t-s)\rho^{\text S}(s){\text d}s,
\end{gather}
where $U_m(\tau) = \exp{(-i(e_m + H_{\text B} + \lambda_{mm}H_{mm})\tau)}$ and
\be
F_{mn}(\tau)=\frac{1}{d}\text{Tr}_{\text B}\left\{ U_{m}^{\dagger}(\tau)H_{mn}U_{n}(\tau)H_{mn}^{\dagger}\right\}.
\ee
Based on the definition of $F_{mn}(t)$, it is easy to see that
\be
F_{mn}(\tau)=F_{nm}^{*}(\tau)=F_{nm}(-\tau).
\ee
Expanding $\rho^{\text S}(s)$ as
\be
\rho^{\text S}(s)=\sum_{m,n=1}^{2}\rho_{mn}^{\text S}(s)|m\rangle\langle n|,
\ee
and rewriting $S_1$ in a more concrete form, one gets 
\be\label{eq-S1}
S_{1}=-|\lambda_{12}|^{2}\left[\begin{array}{cc}
\int_{0}^{t}F_{12}(t-s)\rho_{11}^{\text S}(s)\text{d}s & \int_{0}^{t}F_{12}(t-s)\rho_{12}^{\text S}(s)\text{d}s\\
\int_{0}^{t}F_{21}(t-s)\rho_{21}^{\text S}(s)\text{d}s & \int_{0}^{t}F_{21}(t-s)\rho_{22}^{\text S}(s)\text{d}s
\end{array}\right],
\ee
where $[S_{1}]_{mn}=\langle m|S_{1}|n\rangle$.
Similarly, one has
\be\label{eq-S2}
S_{2}=|\lambda_{12}|^{2}\left[\begin{array}{cc}
\int_{0}^{t}F_{12}(t-s)\rho_{22}^{\text S}(s)\text{d}s & \int_{0}^{t}G_{12}(t,s)\rho_{21}^{\text S}(s)\text{d}s\\
\int_{0}^{t}G_{21}(t,s)\rho_{12}^{\text S}(s)\text{d}s & \int_{0}^{t}F_{21}(t-s)\rho_{11}^{\text S}(s)\text{d}s
\end{array}\right],
\ee
where
\be\label{eq-Gmn}
G_{mn}(t,s)=\frac{1}{d}\text{Tr}_{\text B}\left\{ U_{m}^{\dagger}(t)H_{mn}U_{n}(t)U_{m}^{\dagger}(s)H_{mn}U_{n}(s)\right\} .
\ee
Before moving forward, one need to estimate the correlation function $G_{mn}(t,s)$, where we consider $G_{21}(t,s)$ as an example, which can be written as
\be
G_{21}(t,s)=\frac{1}{d}\text{Tr}_{\text B}\left\{ U_{12}(s,t)H_{21}U_{12}(t,s)H_{21}\right\},
\ee
where 
\be
U_{mn}(s,t) \equiv U_{m}(s)U_{n}^{\dagger}(t).
\ee
Denoting the eigenstate of $U_{12}(s,t)$ and $U_{12}(t,s)$ by $|k\rangle$ and $|k^\prime\rangle$ respectively, $G_{21}(t,s)$ can be further rewritten as
\begin{gather}
G_{21}(t,s)= \frac{1}{d} \sum_{kk^\prime}\langle k|U_{12}(s,t)|k\rangle\langle k^{\prime}|U_{12}(s,t)|k^{\prime}\rangle \nonumber \\
 \cdot [H_{21}]_{kk^{\prime}}[H_{21}]_{k^{\prime}k},
\end{gather}
where
\be
[H_{21}]_{kk^{\prime}} \equiv \langle k | H_{21} |k^\prime\rangle
\ee
As $H_{21}$ is not Hermitian, so in general cases $[H_{21}]_{kk^{\prime}}$ and $[H_{21}]_{k^{\prime}k}$ don't have strong correlations, thus $R_{kk^\prime}\equiv[H_{21}]_{kk^{\prime}}[H_{21}]_{k^{\prime}k}$ can be taken as random numbers with mean zero and variance $\frac{1}{d^2}$ for $k\neq k^\prime$. At the same time, the diagonal elements
$R_{kk}$ scale as $R_{kk} \sim \frac{1}{d}$.
Combining with the fact that the diagonal elements of $U_{12}(t,s)$ and $U_{12}(s,t)$ are of order $1$, one has the following estimation for $G_{21}(t,s)$
\be
G_{21}(t,s)\sim \frac{1}{d}( \sum_{k}R_{kk}+\sum_{\substack{k,k^{\prime}\\
k\neq k^{\prime}
}
}R_{kk^{\prime}} ) \sim\frac{1}{d}.
\ee
Similarly one has 
\be
G_{12}(t, s) \sim \frac{1}{d}.
\ee
If the Hilbert space dimension of the bath is sufficient large, the off-diagonal part of $S_2$ in Eq. \eqref{eq-S2} can be neglected, which yields
\be\label{eq-S2-New}
S_{2}=|\lambda_{12}|^{2}\left[\begin{array}{cc}
\int_{0}^{t}F_{12}(t-s)\rho_{22}^{\text S}(s)\text{d}s & 0\\
0 & \int_{0}^{t}F_{21}(t-s)\rho_{11}^{\text S}(s)\text{d}s
\end{array}\right].
\ee
Inserting Eqs. \eqref{eq-S1} and \eqref{eq-S2-New} to Eq. \eqref{eq-drhoS}, one has, for the diagonal elements
\begin{align}
\frac{{\text{d}}\rho_{11}^{\text S}(t)}{{\text{d}}t}&=-4|\lambda_{12}|^{2}\int_{0}^{t}\mathrm{Re}[F_{12}(t-s)](\rho_{11}^{\text S}(s)-\rho_{11}^{\text S}(\text{eq})){\text{d}}s, \nonumber \\
\frac{\text{d}\rho_{22}^{{\text S}}(t)}{\text{d}t}&=-\frac{\text{d}\rho_{11}^{{\text S}}(t)}{\text{d}t}, 
\end{align}
as well as for off-diagonal elements,
\be
\frac{{\text d}\rho_{21}^{\text S}(t)}{{\text d}t}=-2|\lambda_{12}|^{2}\int_{0}^{t}\mathrm{Re}[F_{12}(t-s)]\rho_{21}^{\text S}(s){\text d}s,
\ee
where
\be
\rho_{11}^{{\text S}}(\text{eq})=\rho_{22}^{{\text S}}(\text{eq})=\frac{1}{2}.
\ee
Under the condition that the the correlation function $F_{12}(\tau)$ decay sufficiently fast on a time $\tau_B$ (correlation time) which is small compare to the relaxation time of the system $\tau_R$, that is, 
\be\label{eq-condition}
\tau_B \ll \tau_R,
\ee
one can employ the Markov approximation.
As a result,  one obtains Eq. \eqref{eq-dRho-both}, which is the Markovian master equation in the interaction picture. 

\section{Generalization of Eq. \eqref{eq-F12-00} to a $N$-level system}\label{sect-mlevel}
In this section, we discuss the relation between the Loschmidt echo $L_{mn}^\text{typ}(t)$ (Eq. \eqref{eq-asm-dec2}) and bath correlation function $F_{mn}(t)$ (Eq. \eqref{def-Fmn}) in a more general setup, 
where a $N-$level system is coupled to a bath. The Hamiltonian reads,
\begin{gather}
H=H_{\text S}+H_{\text B}+V, \nonumber \\ 
V = \sum_{m,n=1}^{N}\lambda_{mn}|m\rangle\langle n|\otimes H_{mn}.
\end{gather}
We employ the same assumption that $H_{mn}$ are uncorrelated with each other.
Eigenstates of $H_{\text S}$ and $H_{\text B}$ are denoted by $|m\rangle$ and $|E_k\rangle$, 
\be
H_{\text S} |m\rangle = e_m |m\rangle,\quad H_{\text B}|E_k\rangle = E_k |E_k\rangle.
\ee
Employing the modified Redfield theory, the unperturbed Hamiltonian and the perturbation are written as
\begin{gather}
H_{0}=H_{\text{S}}+H_{\text{B}}+\sum_{m=1}^{N}|m\rangle\langle m|\otimes H_{mm}, \nonumber \\
H^\text{int} = \sum_{\substack{m,n=1\\
m\neq n
}
}^{N}|m\rangle\langle n|\otimes H_{mn}.
\end{gather}
The bath correlation function reads
\be
F_{mn}(\tau)=\frac{1}{d}\text{Tr}_{B}\{U_{m}^{\dagger}(\tau)H_{mn}U_{n}(\tau)H_{mn}^{\dagger}\},
\ee
where
\be
U_{m}(\tau)=e^{-i(e_m+H_{\text{B}}+\lambda_{mm}H_{mm})t}.
\ee
In case of $\lambda_{mm} \gg 1$, the LE term $L^\text{typ}_{mn}(t)$ (which characterizes decoherence process induced by EC interaction) can be written as
\be
L_{mn}^{\text{typ}}(t)\simeq\frac{1}{d}\text{Tr}_{\text{B}}\{e^{i\lambda_{mm}H_{mm}t}e^{-i\lambda_{nn}H_{nn}t}\}.
\ee
If $H_{mn}$ are uncorrelated, following similar derivations as in Sec. \ref{sect-RDMS} and \ref{sect-relation}, one gets that
\begin{gather}
F_{mn}(t)\simeq\frac{1}{d^{2}}\text{Tr}_{\text{B}}\{H_{mn}H_{mn}^{\dagger}\}\frac{1}{d}\text{Tr}_{\text{B}}\{e^{i\lambda_{mm}H_{mm}t}\}\text{Tr}_{\text{B}}\{e^{-i\lambda_{nn}H_{nn}t}\} \nonumber \\
L_{mn}^{\text{typ}}(t)\simeq\frac{1}{d^{2}}\text{Tr}\{e^{i\lambda_{mm}H_{mm}t}\}\text{Tr}\{e^{-i\lambda_{nn}H_{nn}t}\}. \label{eq-mlevel}
\end{gather}
From Eq. \eqref{eq-mlevel} one can see that
\be
F_{mn}(t)\simeq\frac{1}{d^{2}}\text{Tr}_{\text{B}}\{H_{mn}H_{mn}^{\dagger}\}L_{mn}^{\text{typ}}(t),
\ee
which is a generalization of Eq. \eqref{eq-F12-00} to a $N$-level system.
\end{appendix}
 
\end{document}